\documentclass[pra,twocolumn,floatfix,a4paper,superscriptaddress]{revtex4}
\usepackage{bm,color,graphicx,amsmath,txfonts}
\usepackage[colorlinks=true, allcolors=blue]{hyperref} 

\newcommand{\CC}{{\cal C}}

\begin{document}

\title{Stationary quantum entanglement between a massive mechanical membrane\\ and a low frequency LC circuit}\thanks{This work was published in \href{https://doi.org/10.1088/1367-2630/ab90d2}{New J.\ Phys.\ \textbf{22}, 063041} (2020). The source data for the figures is available at \href{https://doi.org/10.5281/zenodo.3906605}{10.5281/zenodo.3906605}.}

\author{Jie Li}\thanks{jieli6677@hotmail.com}
\affiliation{Kavli Institute of Nanoscience, Department of Quantum Nanoscience, Delft University of Technology, 2628CJ Delft, The Netherlands}
\author{Simon Gr\"oblacher}\thanks{s.groeblacher@tudelft.nl}
\affiliation{Kavli Institute of Nanoscience, Department of Quantum Nanoscience, Delft University of Technology, 2628CJ Delft, The Netherlands}

\begin{abstract}
We study electro-mechanical entanglement in a system where a massive membrane is capacitively coupled to a {\it low frequency} LC resonator. In opto- and electro-mechanics, the entanglement between a megahertz (MHz) mechanical resonator and a gigahertz (GHz) microwave LC resonator has been widely and well explored, and recently experimentally demonstrated. Typically, coupling is realized through a radiation pressure-like interaction, and entanglement is generated by adopting an appropriate microwave drive. Through this approach it is however not evident how to create entanglement in the case where both the mechanical and LC oscillators are of low frequency, e.g., around 1 MHz. Here we provide an effective approach to entangling two low-frequency resonators by further coupling the membrane to an optical cavity. The cavity is strongly driven by a red-detuned laser, sequentially cooling the mechanical and electrical modes, which results in stationary electro-mechanical entanglement at experimentally achievable temperatures. The entanglement directly originates from the electro-mechanical coupling itself and due to its quantum nature will allow testing quantum theories at a more macroscopic scale than currently possible.
\end{abstract}

\date{\today}
\maketitle

\section{Introduction}

In optomechanics, an optical field can couple to a massive mechanical oscillator (MO) via the radiation pressure force~\cite{omRMP}. This approach provides the possibility to prepare quantum states of macroscopic systems by manipulating optical degrees of freedom. Over the past decade, significant experimental progress has been achieved in observing quantum effects in massive mechanical systems, including reaching the quantum ground state~\cite{ground1,ground2}, quantum squeezing of the mechanical motion~\cite{sqz1,sqz2,sqz3}, quantum entanglement between two MOs~\cite{enMM1,enMM2}, and between a MO and an electromagnetic field~\cite{enOM1,enOM2}, among many others. Such quantum states of massive objects have important implications for both quantum technologies, e.g., quantum sensing~\cite{sensing}, quantum transducers~\cite{SG}, as well as foundational studies of decoherence theories at the macro scale and the boundary between the quantum and classical worlds~\cite{review}.

In this paper, we provide a scheme to entangle a massive MO with a macroscopic low-frequency LC oscillator. Specifically, we consider a tripartite system where a mechanical membrane is capacitively coupled to an LC resonator and further optomechanically coupled to an optical cavity. Unlike most other approaches with GHz resonators~\cite{ground1,enOM1,Lehnert,Fink}, the LC resonator we consider here is in the {\it radio frequency} domain, around 1 MHz~\cite{Polzik14,Iman18}, and close to the mechanical frequency. Such a low-frequency LC resonator means a much larger product $L\times C$ ($L$-inductance; $C$-capacitance) than that at microwave frequency ($10^6$ larger for frequency of 1~MHz compared to 1~GHz), which typically implies a much larger number of charges and a much bigger LC circuit. The membrane-LC interaction takes a nonlinear form $H_{int} = \hbar g_0 x q^2$~\cite{DV07,DV11,Polzik11}, where $x$ is the mechanical position, $q$ the charge and $g_0$ the bare electro-mechanical coupling rate, which is a radiation pressure-like interaction. We note that the entanglement between a nanomechanical resonator and an LC resonator of {\it microwave frequency} has been well studied~\cite{DV07,DV11}, where the electro-mechanical interaction is a radiation pressure type $\propto g_0 x b^{\dag}b$. Here $b$ is the annihilation operator of the LC field and $q=(b+b^{\dag})/\sqrt{2}$. The interaction is derived by taking the rotating wave approximation (RWA) and neglecting fast oscillating terms, which is valid only for the LC frequency much larger than the mechanical frequency, $\omega_{LC} \gg \omega_m$. Such a radiation pressure interaction predicts the generation of electro-mechanical entanglement if an appropriate microwave drive is adopted~\cite{DV07}. However, when the LC frequency is approaching the mechanical frequency, like in Ref.~\cite{Polzik14,Iman18} and as considered here, it is not clear how to apply an appropriate driving field such that (stationary) electro-mechanical entanglement can be produced. In other words, it is not apparent how to apply the mechanism of Ref.~\cite{DV07} to two nearly resonant low-frequency oscillators.

Inspired by recent experiments~\cite{Polzik14,Iman18}, we apply a DC drive for the LC circuit, which significantly enhances the effective electro-mechanical coupling rate. The linearized interaction takes the form $\propto g \delta x \delta q$~\cite{Polzik11,Polzik14}, where $g$ is the effective coupling rate. Based on this ``quadrature-quadrature" coupling, electro-mechanical entanglement can indeed be created, but only at unrealistic extremely low temperature (below 0.1 mK for 1 MHz oscillators), where both the oscillators are actually at their quantum ground state. In this situation, the component of the beamsplitter interaction in the ``quadrature-quadrature" coupling is significantly suppressed, while the component of the two-mode squeezing interaction plays a dominate role, leading to the generation of electro-mechanical entanglement. The entanglement completely vanishes at typical cryogenic temperatures of a few tens of millikelvin, because of the low resonant frequencies and thus large thermal occupations. In our approach, we overcome this limitation and show that by further coupling the mechanical membrane to an optical cavity field via radiation pressure, and by driving the cavity with a red-detuned laser, both the mechanical and electrical modes get significantly cooled, which leads to the emergence of electro-mechanical entanglement. Here, the red-detuned cavity cools the mechanical mode, which then acts as a cold bath for the electrical mode~\cite{LCcooling}. The entanglement is in the steady state regime and robust against temperature.  

The remainder of the paper is organized as follows:\ in Sec.~\ref{model}, we introduce our tripartite opto-electro-mechanical system, provide its Hamiltonian and the corresponding Langevin equations, and in Sec.~\ref{Solu} we show how to obtain the steady-state solutions of the system and quantify the entanglement. In Sec.~\ref{Ent}, we present the results of electro-mechanical entanglement and discuss optimal parameter regimes for obtaining the entanglement and its detection. Finally, we draw the conclusions in Sec.~\ref{conc}.

\begin{figure}[t]
\hskip-0.4cm\includegraphics[width=0.95\linewidth]{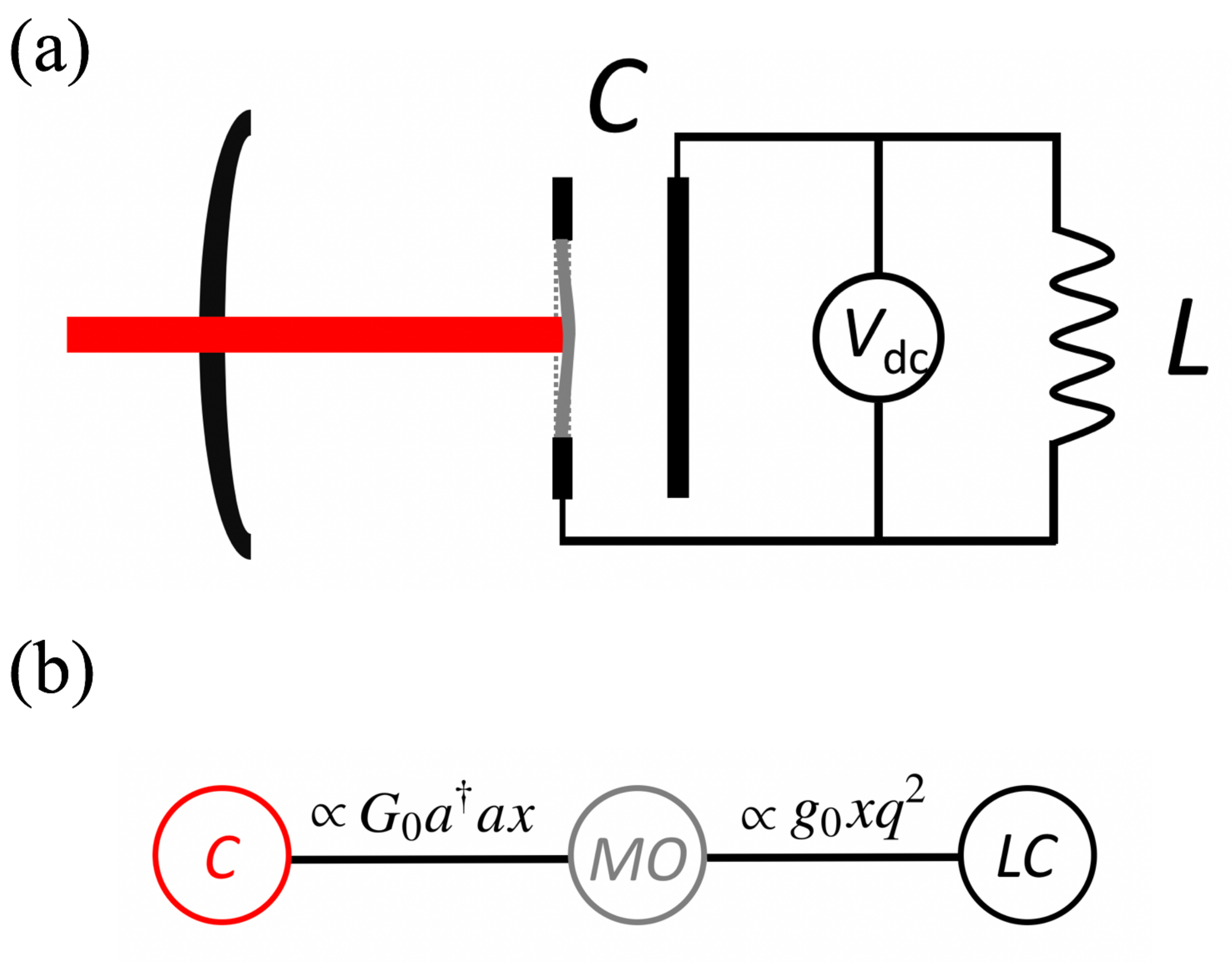} 
\caption{(a) A mechanical oscillator, embodied by a metal coated membrane, is capacitively coupled to an LC circuit and simultaneously coupled to an optical cavity via radiation pressure, where the membrane forms one end mirror of the cavity. (b) The mechanical oscillator couples to the cavity via nonlinear radiation pressure interaction $\propto G_0 a^{\dag} a x$, and to the LC circuit via radiation pressure-like interaction $\propto g_0 x q^2$. The frequencies we consider for the mechanical and LC oscillators are around 1 MHz, much lower than the cavity resonant frequency.}
\label{fig1}
\end{figure}

\section{The system}
\label{model}

We consider a tripartite opto-electro-mechanical system, as shown in Fig.~\ref{fig1}, which consists of an LC electrical circuit, a MO, and an optical cavity. An experimental realization of a suitable MO could be a metal coated nanomembrane~\cite{Polzik14,Iman18}, which is capacitively coupled to an LC resonator and further coupled to an optical cavity field via radiation pressure. Specifically, the radiation pressure of the cavity field causes a mechanical displacement which further changes the capacitance of the LC circuit, and conversely, the voltage fluctuation in the LC circuit leads to an optical phase shift via the mediation of the MO. The Hamiltonian of the system reads
\begin{equation}
\begin{split}
H/\hbar = & \omega_c a^{\dag} a + \frac{\omega_m}{2} (x^2 + p^2) + \frac{\omega_{LC}}{2} (q^2 + {\phi}^2) - G_0 a^{\dag} a x \\
&+ g_0 x q^2  - q \frac{q_0}{\hbar} V + i E (a^{\dag} e^{-i \omega_l t}  - a e^{i \omega_l t} ),
\end{split}
\end{equation}
where $a$ ($a^{\dag}$) is the annihilation (creation) operator of the cavity mode, $x$ and $p$ ($q$ and $\phi$) are the {\it dimensionless} position and momentum (charge and flux) quadratures of the mechanical (LC) resonator, and therefore $[a, a^{\dag}]\,{=}\,1$ and $[x, p]\,{=}\,[q, \phi]\,{=}\,i$. The resonance frequencies $ \omega_c$, $ \omega_m$, and $ \omega_{LC}\,{=}\,\frac{1}{\sqrt{LC}}$ are of the cavity, mechanical, and LC resonators, respectively, where $L$ ($C$) is the inductance (capacitance) of the LC circuit. The capacitance $C(x)$ is a function of the mechanical position $x$, which characterizes the capacitive coupling to the MO. The MO-LC nonlinear coupling $\propto g_0 x q^2$ is derived by expanding the capacitance $C(x)$ around the membrane equilibrium position and expanding the capacitive energy $\frac{q^2}{2C(x)}$ as a Taylor series up to first order~\cite{DV07,Polzik11}, and $G_0$ is the single-photon optomechanical coupling rate. The last two terms in the Hamiltonian denote the electric driving for the LC circuit and the laser driving for the cavity, respectively, where $V$ is a DC bias voltage (see Fig.~\ref{fig1} (a)), $q_0\,{=}\,\sqrt{\hbar/L\omega_{LC}}$ is the zero-point fluctuation of the LC oscillator, and $E=\!\sqrt{2P_l \kappa/\hbar \omega_l}$ is the coupling between the cavity field with decay rate $\kappa$ and the driving laser with frequency $\omega_l$ and power $P_l$. Note that Refs.~\cite{DV07,DV11} considered a different form of MO-LC coupling, $\propto g_0 x b^{\dag}b$, which was derived by neglecting fast oscillating terms $xbb$ and $xb^{\dag}b^{\dag}$ valid only when $\omega_{LC} \gg \omega_m$. Instead, here we consider two low frequency nearly resonant oscillators $\omega_{LC} \simeq \omega_m$, and thus one has to consider the full interaction $\propto g_0 x q^2$. Another major difference is that the blue- or red-detuned pump used in Refs.~\cite{DV07,DV11} to activate {\it electro-mechanical} Stokes or anti-Stokes process does not apply, in a clear way, to our case of nearly resonant oscillators.

In the frame rotating at the drive frequency $\omega_l$, the quantum Langevin equations (QLEs) governing the system dynamics are given by
\begin{equation}\label{QLE1}
\begin{split}
\dot{a}&= - (i \Delta_0 + \kappa) a + i G_0 x a + E + \sqrt{2 \kappa} a^{\rm in},  \\
\dot{x}&= \omega_m p,   \\
\dot{p}&= - \omega_m x - \gamma_m p + G_0 a^{\dag} a - g_0 q^2  + \xi,   \\
\dot{q}&= \omega_{LC} \phi,   \\
\dot{\phi}&= - \omega_{LC} q - \gamma_{LC} \phi - 2 g_0 x q  +\frac{q_0}{\hbar} V,   \\
\end{split}
\end{equation}
where $\Delta_{0} \,\,{=}\,\, \omega_{c} \,{-}\, \omega_l$, $\gamma_m$ and $\gamma_{LC} \,\,{=}\,\, 2R/L$ (with $R$ the resistance of the circuit) are the mechanical and electrical damping rates, respectively, $a^{\rm in}$ is the input noise operator for the cavity, whose mean value is zero and the only non-zero correlation is 
\begin{equation}
\langle a^{\rm in}(t) \, a^{\rm in \dag}(t')\rangle = \delta(t-t').
\end{equation} 
The Langevin force operator $\xi$ accounts for the Brownian motion of the MO and is autocorrelated as 
\begin{equation}
\langle \xi(t)\xi(t')+\xi(t') \xi(t) \rangle/2  \simeq \gamma_m (2 \bar{n}_m+1) \delta(t-t'), 
\end{equation}
where we have made a Markovian approximation valid for large mechanical quality factors $Q_m= \omega_m/\gamma_m \,\, {\gg}\, 1$~\cite{Markov}, and $\bar{n}_m \simeq \frac{k_B T}{\hbar \omega_m}$ is the equilibrium mean thermal phonon number in the high temperature limit, with $k_B$ the Boltzmann constant and $T$ the environmental temperature.

In order to get strong optomechanical (electro-mechanical) coupling for cooling both the mechanical and LC oscillators (creating MO-LC entanglement), we consider an intense laser pump, leading to a large amplitude of the cavity field $|\langle a\rangle| \gg 1$, and a large number of charges $\langle q\rangle \gg 1$. This allows us to linearize the system dynamics around the semiclassical averages by writing any operator as $O=\langle O \rangle +\delta O$ ($O\, {=}\, a,x,p,q,\phi$) and neglecting small second-order fluctuation terms. Therefore, the QLEs Eq.~\eqref{QLE1} are separated into two sets of equations: one is for averages $O_s \equiv \langle O \rangle$ and the other for zero-mean quantum fluctuations $\delta O$. The steady-state averages can be obtained by setting the derivatives to zero and solving the following equations
\begin{equation}\label{Ave}
\begin{split}
a_s &=  \frac{E}{\kappa+ i \Delta}  ,\\
x_s &= \frac{1}{\omega_m} \Big( G_0 |a_s|^2 - g_0 q_s^2  \Big)  ,\\
q_s &= \frac{1}{\omega'_{LC}} \Big(  \frac{q_0}{\hbar}  \bar{V} \Big)  ,\\
p_s &=  \phi_s = 0,   \\ 
\end{split}
\end{equation}
where $\Delta \,\,{=}\,\, \Delta_0 \,{-}\, G_0 x_s$ is the effective cavity-laser detuning, and $\omega'_{LC}=\omega_{LC}+2g_0x_s$ is the effective LC frequency including the frequency shift caused by the nonlinear MO-LC interaction. The linearized QLEs describing the quadrature fluctuations $(\delta X, \delta Y, \delta x, \delta p, \delta q, \delta \phi)$, with $\delta X=(\delta a + \delta a^{\dag})/\sqrt{2}$, $\delta Y=i(\delta a^{\dag} - \delta a)/\sqrt{2}$, are given by
\begin{equation}\label{QLE2}
\begin{split}
\delta \dot{X}&= \Delta \delta Y - \kappa \delta X + \sqrt{2 \kappa} X^{\rm in} ,  \\
\delta \dot{Y}&= - \Delta \delta X - \kappa \delta Y + G \delta x + \sqrt{2 \kappa} Y^{\rm in} ,  \\
\delta \dot{x}&= \omega_m \delta p,   \\
\delta \dot{p}&= - \omega_m \delta x - \gamma_m \delta p + G \delta X - g \delta q  + \xi,   \\
\delta \dot{q}&= \omega'_{LC} \delta \phi,   \\
\delta \dot{\phi}&= - \omega'_{LC} \delta q - \gamma_{LC} \delta \phi - g \delta x  +\frac{q_0}{\hbar} \delta V,   \\
\end{split}
\end{equation}
where $G= \! \sqrt{2} G_0  a_s$ ($g=2 g_0 q_s$) is the effective optomechanical (electro-mechanical) coupling rate, and $X^{\rm in}=(a^{\rm in} + a^{\rm in \dag})/\sqrt{2}$, $Y^{\rm in}=i(a^{\rm in \dag} - a^{\rm in})/\sqrt{2}$ are the quadratures of the cavity input noise. Note that in deriving the above QLEs, we have chosen a phase reference such that $a_s$ is real and positive. The effective coupling $g$ increases linearly with $q_s$, which then is linear dependence on the bias voltage $\bar{V}$ (see Eq.~\eqref{Ave}). This means that the electro-mechanical coupling strength can be significantly improved by increasing the bias voltage~\cite{Polzik14}. The fluctuation of the bias voltage $\delta V\, {\equiv}\, V \,{-} \,\bar{V}$ can be considered as the input noise for the flux, and is autocorrelated as
\begin{equation} 
\langle \delta V(t) \delta V(t') \rangle = \bigg[4 k_B T R+ \gamma_{LC} \Big( \frac{\hbar}{q_0} \Big)^2 \bigg] \, \delta(t-t'),
\end{equation} 
which corresponds to the quantum version of the Johnson-Nyquist noise correlation~\cite{JN} for a resistor $R=\frac{\gamma_{LC}}{2} L$ at temperature $T$ by including the vacuum fluctuation. In such a way, the noise correlation for the operator $\delta {\cal V} (t)  \equiv \frac{q_0}{\hbar} \delta V (t)$ can be written in the form 
\begin{equation} 
\langle \delta {\cal V}(t) \delta {\cal V}(t') \rangle = \gamma_{LC} (2 \bar{n}_{LC}+ 1) \, \delta(t-t'), 
\end{equation} 
with $\bar{n}_{LC} \simeq \frac{k_B T}{\hbar \omega_{LC}}$ being the thermal occupancy of the LC oscillator, which takes a consistent form as that for the Langevin force operator $\xi$. This is the reason why we defined the damping rate $\gamma_{LC}$ as twice its conventional definition $\gamma'_{LC} = R/L$.

\section{Steady-state solutions and quantification of Gaussian entanglement}
\label{Solu}

We are interested in the quantum correlation between the mechanical and LC oscillators in the stationary state. Owing to the fact that the dynamics are linearized and all input noises are Gaussian, the Gaussian nature of the state will be preserved for all times. The steady state of the quantum fluctuations of the system is therefore a three-mode Gaussian state and is completely characterized by a $6\times6$ covariance matrix (CM) $\CC$, which is defined as $\CC_{ij} \,{=}\, \frac{1}{2}\langle u_i(t) u_j(t') + u_j(t') u_i(t)   \rangle$ ($i,j \,{=}\, 1,2,...,6$), where $u(t)=\big[\delta X (t), \delta Y (t), \delta x (t), \delta p (t), \delta q (t), \delta \phi (t) \big]^T$. The stationary CM $\CC$ can be obtained by solving the Lyapunov equation~\cite{Hahn}
\begin{equation}\label{Lyap}
A \CC + \CC A^T = -D,
\end{equation}
where $A$ is the drift matrix determined by the QLEs~\eqref{QLE2}, given by  
\begin{equation}\label{AAA}
A =
\begin{pmatrix}
-\kappa  &  \Delta  &  0 &  0  &  0  &  0   \\
-\Delta  & -\kappa  & G  & 0  &  0  &  0   \\
0 & 0 & 0 & \omega_m  &  0 &  0 \\
G  & 0 & -\omega_m & -\gamma_m &  -g  &  0 \\
0 &  0  &  0  &  0  &  0  &  \omega'_{LC}   \\
0 &  0  &  -g  &  0  & -\omega'_{LC} & -\gamma_{LC}   \\
\end{pmatrix} ,
\end{equation}
and $D = {\rm diag} \big[ \kappa, \, \kappa, \, 0, \, \gamma_m (2\bar{n}_m +1 ), \, 0,  \, \gamma_{LC} (2\bar{n}_{LC} +1 ) \big]$ is the diffusion matrix, which is defined by $\langle  n_i(t) n_j(t') +n_j(t') n_i(t) \rangle/2 = D_{ij} \delta (t-t')$, with the vector of input noises $n (t) \,{=}\, \big[ \!\sqrt{2\kappa} X^{\rm in} (t), \! \sqrt{2\kappa} Y^{\rm in} (t), 0, \xi (t), 0, \delta {\cal V} (t) \big]^T$. To quantify the Gaussian entanglement, we adopt the logarithmic negativity~\cite{LogNeg}, which is a full entanglement monotone under local operations and classical communication~\cite{Plenio} and sets an upper bound for the distillable entanglement~\cite{LogNeg}. The logarithmic negativity is defined as~\cite{Adesso}
\begin{equation}\label{LogNeg}
E_N \equiv \max[0, \, -\ln2\tilde\nu_-],
\end{equation}
where $\tilde\nu_-\,\,{=}\,\min{\rm eig}|i\Omega_2\tilde{\CC}_4|$ (with the symplectic matrix $\Omega_2=\oplus^2_{j=1} \! i\sigma_y$ and the $y$-Pauli matrix $\sigma_y$) is the minimum symplectic eigenvalue of the partially transposed CM $\tilde{\CC}_4={\cal P}_{1|2}{\CC_4}{\cal P}_{1|2}$, with $\CC_4$ being the $4\times 4$ CM of the mechanical and electrical modes, obtained by removing in $\CC$ the rows and columns related to the cavity field, and ${\cal P}_{1|2}={\rm diag}(1,-1,1,1)$ being the matrix that performs partial transposition on CM~\cite{Simon}.

\begin{figure}[b]
\hskip0cm\includegraphics[width=\linewidth]{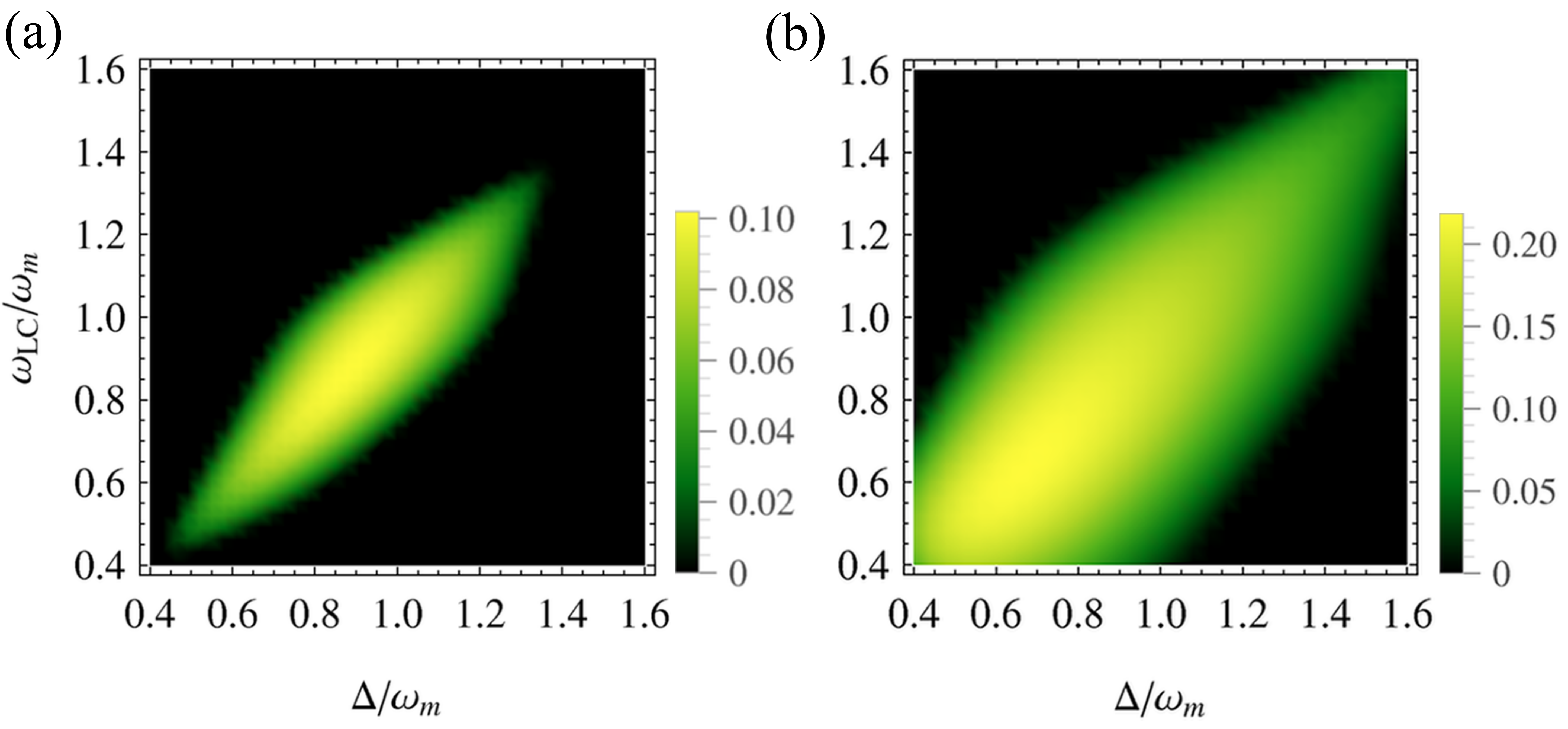} 
\caption{Electro-mechanical entanglement $E_N$ vs.\ detuning $\Delta/\omega_m$ and $\omega_{LC}/\omega_m$ (with $\omega_m$ fixed). We take $G=g=3\kappa$ in (a) and $5\kappa$ in (b), while the surrounding bath temperature $T=10$~mK. See text for the other parameters.}
\label{fig2}
\end{figure}

\section{Electro-mechanical entanglement in the steady state}
\label{Ent}

In this section, we present the results of the entanglement between the mechanical and LC oscillators. All results are in the steady state guaranteed by the negative eigenvalues (real parts) of the drift matrix $A$. We adopt experimentally feasible parameters~\cite{Polzik14,Iman18}: $\omega_c/2\pi = 200$ THz, $\omega_m/2\pi = 1$ MHz, $\kappa=0.1 \omega_m$, $\gamma_m= 10^{-6} \omega_m$, $\gamma_{LC}= 10^{-5} \omega_{LC}$, and consider the LC frequency $\omega_{LC}$ as a variable which is tuned around $\omega_m$. To avoid additional low-frequency electronic noises, LC frequencies much below 1 MHz will not be considered. We work in the resolved sideband limit, $\kappa \ll \omega_m$, and assume a relatively large $Q$ factor of the LC oscillator compared to those typically demonstrated at room temperature~\cite{Polzik14,Iman18} as we place the system at cryogenic temperatures where superconductivity can significantly improve the $Q$ factor~\cite{Gary}. At a few tens of millikelvin, the mechanical and LC oscillators still exhibit significant thermal excitations because of their low frequencies. Therefore, we use a red-detuned laser to drive the cavity and stimulate the optomechanical anti-Stokes process, which results in cooling of the mechanical mode~\cite{ground2}, and owing to the MO-LC coupling, the electrical mode also gets cooled. In such a system, it is even possible to cool a 1 MHz LC resonator into its quantum ground state from temperature of a few tens of millikelvin~\cite{LCcooling}. The cooling process in this hybrid system can be considered as the transport of thermal excitations from the electrical mode to the mechanical mode, and then to the cavity mode, which eventually dissipates the heat via cavity photon leakage to the environment. The low effective temperatures of the mechanical and electrical modes are a precondition for observing their entanglement if strong coupling rates are used. This is verified numerically and shown in Fig.~\ref{fig2}, where the entanglement is maximal for a cavity-laser detuning $\Delta \simeq \omega_{LC}$. We assume both the optomechanical and the electro-mechanical coupling to be strong $G, g>\kappa$, in order to significantly cool both the mechanical and electrical modes~\cite{LCcooling} and to ultimately create the desired electro-mechanical entanglement. We have verified that based on the values of the bare coupling rates $G_0$ and $g_0$ estimated from the experiments~\cite{Lehnert,Polzik14,Iman18}, with the parameters used for our results the nonlinear electro-mechanical coupling induced frequency shift $\omega'_{LC} \,{-}\, \omega_{LC} \,{\ll}\, \omega_{LC}$. Therefore, throughout the paper we consider $\omega'_{LC} \,{\simeq}\, \omega_{LC}$. Figure~\ref{fig2} also shows that in our system two nearly resonant oscillators are preferred to maximize the entanglement. If the couplings are further increased (cf.\ Fig.~\ref{fig2}(b)) the system becomes unstable for $\Delta \, {<} \,{\sim} \, 0.4 \omega_m$.

\begin{figure}[t]
\hskip0cm\includegraphics[width=0.95\linewidth]{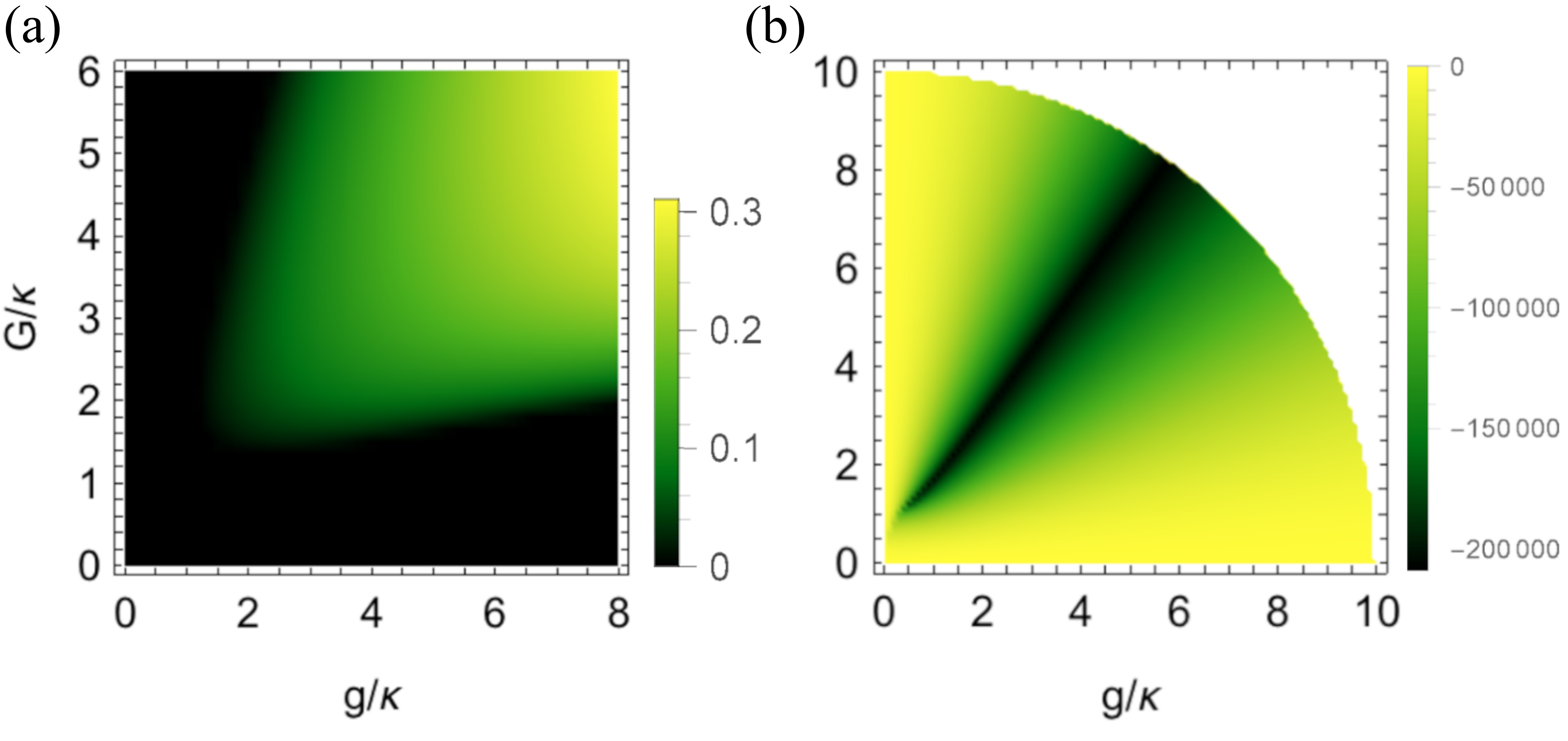} 
\caption{(a) Stationary MO-LC entanglement $E_N$, (b) The maximum of the eigenvalues (real parts) of the drift matrix $A$ vs.\ coupling rates $g$ and $G$. The white area denotes that the system is unstable. We take an optimal detuning $\Delta \simeq \omega_m$ for $\omega_{LC} = \omega_m$, and $T=10$~mK. The other parameters are the same as in Fig.~\ref{fig2}.}
\label{fig3}
\end{figure}

We further show the stationary electro-mechanical entanglement as a function of the two coupling rates $g$ and $G$ for the resonant case~\cite{Polzik14,Iman18} in Fig.~\ref{fig3}(a). It is clear that the entanglement grows with increasing coupling strengths and strong couplings $G, g >\kappa$ are generally required to obtain considerable entanglement. The coupling strengths are restricted by the stability condition. Note that the system becomes stable only when all the eigenvalues of the drift matrix $A$ have negative real parts. The parameter regime where stability occurs can be obtained from the Routh-Hurwitz criterion~\cite{RH}, but the inequalities are quite involved for the present tripartite system. Therefore, to provide an intuitive picture we numerically plot the maximum of the eigenvalues (real parts) of the drift matrix $A$ in Fig.~\ref{fig3}(b). The white area denotes that the maximum is larger than zero, i.e., at least one eigenvalue has a positive real part, and thus the system is unstable when the couplings lie within this area. Under such strong couplings, one may conjecture the ``optical-spring" effect may play a significant role. We therefore derive the expression of the effective mechanical frequency (see Appendix for the derivation), which in the frequency domain is given by
\begin{equation}
\begin{split}
&\omega_m^{\rm eff}(\omega) =  \\
& \Bigg[ \omega_m^2 - \frac{ G^2 \Delta \omega_m (\Delta^2 + \kappa^2 - \omega^2)}{ (\Delta^2 + \kappa^2 - \omega^2)^2 + 4 \kappa^2 \omega^2  }  - \frac{ g^2 \omega_{LC} \omega_m  (\omega_{LC}^2 - \omega^2 ) }{ (\omega_{LC}^2 -\omega^2 )^2 + \gamma_{LC}^2 \omega^2  }    \Bigg]^{\frac{1}{2}},
\end{split}
\end{equation}
where the first term is the MO's natural frequency, and the second term is the frequency shift caused by the optomechanical interaction, followed by the third term which denotes a further frequency shift due to the electro-mechanical coupling. Under the optimal condition for entanglement, $\omega=\Delta=\omega_m=\omega_{LC} \equiv \omega_0$, and in the resolved sideband limit $\kappa \ll \omega_m$, we obtain 
\begin{equation}
\omega_m^{\rm eff} (\omega_0)  \simeq \sqrt{ \omega_m^2 - G^2/4 }.
\end{equation}
For what we have used the strongest coupling $G=6\kappa$ and $\kappa =0.1 \omega_m$, we have $\omega_m^{\rm eff} (\omega_0)  \simeq 0.95 \omega_m$, implying that the mechanical frequency in most cases remains unchanged. This is mainly due to the fact that we work in the resolved sideband limit~\cite{omRMP}.

We note that the entanglement originates from the component of the two-mode squeezing interaction $\propto g (\delta m \delta b + \delta m^{\dag} \delta b^{\dag} )$ in the ``quadrature-quadrature" coupling $\propto g \delta x \delta q =  g (\delta m + \delta m^{\dag} ) (\delta b + \delta b^{\dag} )/2 $, where $m$ is the annihilation operator of the mechanical mode. This can be verified by the fact that there will be no entanglement for a weak coupling $g \ll \omega_{m/LC}$, which allows one to make the RWA and the interaction essentially becomes a beamsplitter type $\propto g (\delta m \delta b^{\dag} + \delta m^{\dag} \delta b)$. As clearly visible from Fig.~\ref{fig3}(a), this situation of weak coupling $g < \kappa \ll \omega_{m/LC}$ does not produce any entanglement. Apart from a sufficiently large $g$, $G$ should also be strong, $G>\kappa$, in order to efficiently cool both the oscillators. Taking $G=g=5\kappa$ in Fig.~\ref{fig3}(a) for example, we obtain the average excitation number of the two modes: $\bar{n}_m^{\rm eff} = \frac{1}{2} \big( \langle \delta x^2 \rangle + \langle \delta p^2 \rangle -1 \big) \simeq 0.15$; $ \bar{n}_{LC}^{\rm eff} = \frac{1}{2} \big( \langle \delta q^2 \rangle + \langle \delta \phi^2 \rangle -1 \big) \simeq 0.08$, implying both the oscillators are cooled into their quantum ground state. This yields an entanglement $E_N \simeq 0.18$.

\begin{figure}[b]
\hskip-0.35cm\includegraphics[width=0.92\linewidth]{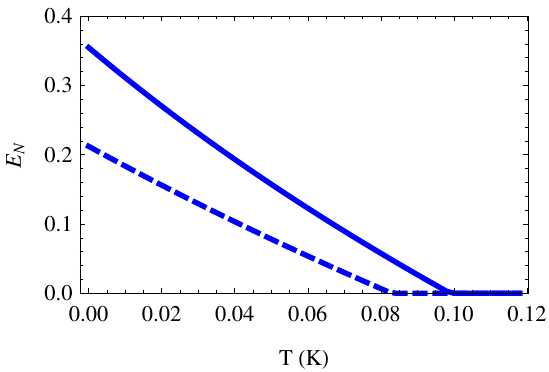} 
\caption{Stationary MO-LC entanglement $E_N$ vs.\ bath temperature $T$:\ solid (dashed) line is for $g=8\kappa$ and $G=6\kappa$ ($g=G=5\kappa$). The other parameters are the same as in Fig.~\ref{fig3}.}
\label{fig4}
\end{figure}

We further investigate the entanglement as a function of bath temperature and the data in Fig.~\ref{fig4} shows that it is robust against temperature, surviving up to $\sim$100~mK, based on realistic parameters. Even though the electro-mechanical coupling rate $g$ and the $Q$ factor of the LC resonator we use are larger than the demonstrated values~\cite{Polzik14,Iman18}, it is realistic to assume that they can be achieved at low temperature and by properly designing the system~\cite{Gary}.

Finally, we would like to discuss how to detect the electro-mechanical entanglement. The task requires to essentially measure the four quadratures of the mechanical and electrical modes, $(x, p, q, \phi)$, based on which the CM can be re-constructed and the logarithmic negativity can then be computed according to the definition in Eq.~\eqref{LogNeg}. To measure the mechanical quadratures, we adopt the strategy used in Refs.~\cite{enOM1,enOM2,DV07L}, i.e., sending a {\it weak} red-detuned probe field with detuning equal to the mechanical frequency $\Delta^p \simeq \omega_m$ into the cavity, which maps the mechanical state onto the anti-Stokes sideband of the probe field at cavity resonance. Thus, by homodyning the probe output field, the two mechanical quadratures are measured. The quadratures of the electrical mode can also be measured by employing a homodyne scheme at radio frequency.

\begin{figure}
\hskip0cm\includegraphics[width=0.95\linewidth]{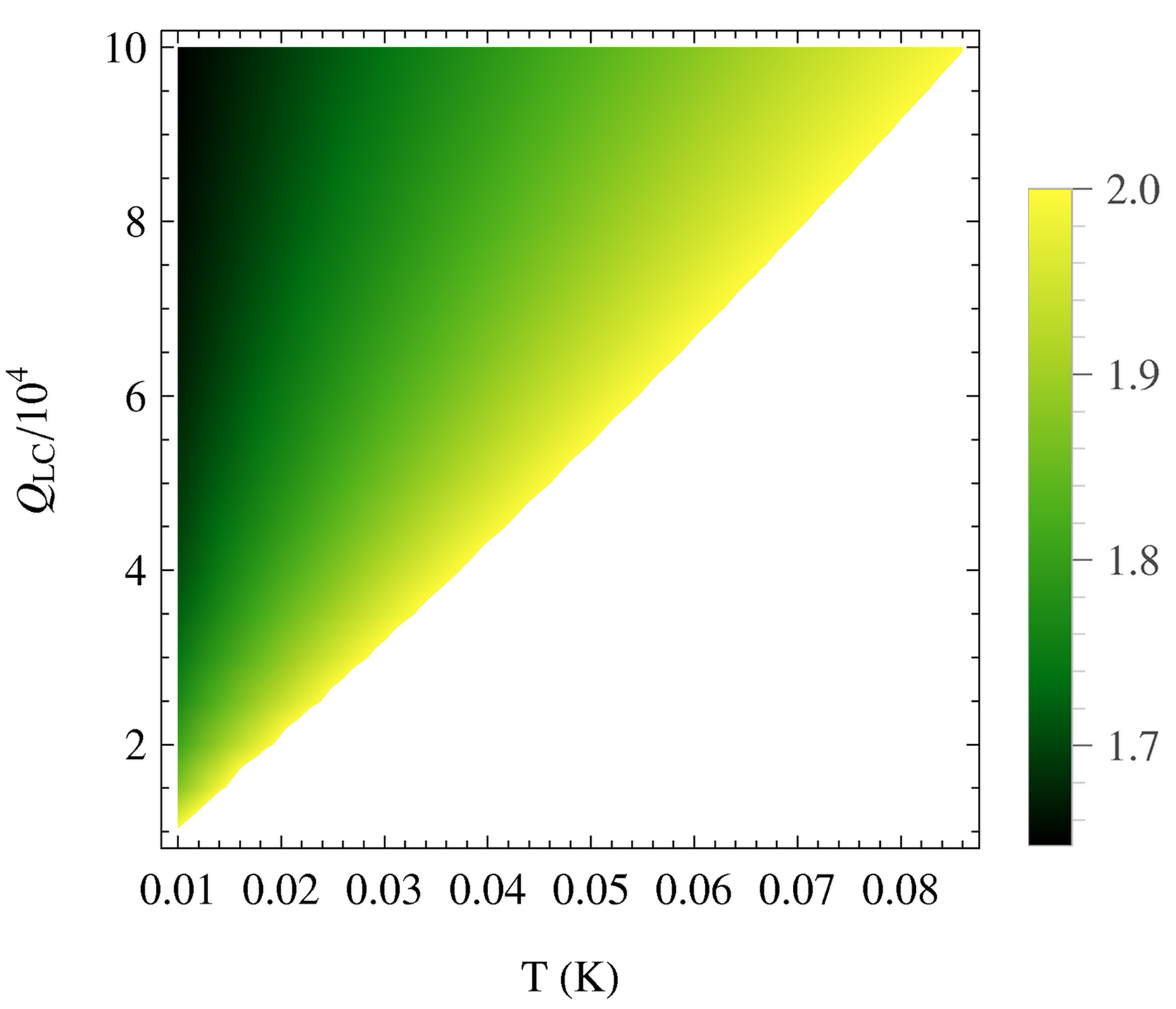} 
\caption{Density plot of $\langle \delta X_+^{2} \rangle + \langle \delta Y_-^{2} \rangle$ vs. $T$ and $Q_{LC}=\omega_{LC}/\gamma_{LC}$ (ranging from $10^4$ to $10^5$ with fixed $\omega_{LC}$) for $g=G=6\kappa$. The white area shows $\langle \delta X_+^{2} \rangle + \langle \delta Y_-^{2} \rangle >2$. The other parameters are the same as in Fig.~\ref{fig3}.}
\label{fig5}
\end{figure}

In order to avoid measuring the whole $4\times 4$ CM for quantifying entanglement, alternatively, one can also {\it verify} the entanglement by using the Duan criterion~\cite{Duan}, which requires the measurement of only two collective quadratures, $X_+ = x +q$, and $Y_- = p - \phi$. A sufficient condition for entanglement is that the two collective quadratures should satisfy the following inequality  
\begin{equation}
\langle \delta X_+^{2} \rangle + \langle \delta Y_-^{2} \rangle < 2.
\end{equation}
Figure~\ref{fig5} shows that in moderate ranges of temperature and LC $Q$ factor the inequality is fulfilled, indicating the presence of electro-mechanical entanglement. The entanglement survives up to 19~mK for $Q_{LC}=2\times 10^4$, and 86~mK for $Q_{LC}=10^5$.

\section{Conclusions} 
\label{conc}

We have provided a straightforward but effective approach to preparing entangled states of low-frequency mechanical and LC resonators. At typical cryogenic temperatures, the two resonators still contain significant thermal excitations, which effectively destroy their joint quantum correlations. In order to solve this, we couple the mechanical element to an optical cavity via the radiation pressure force, which can act as an additional cold bath:\ by drving the cavity with a red-detuned laser, both the mechanical and electrical modes are sequentially cooled, resulting in remarkable electro-mechanical entanglement emerging from thermal noise. The entanglement originates from the electro-mechanical coupling and can be in the stationary state and robust against temperature.

The present work can be considered as a complementary study to the widely explored situation where the LC frequency, typically in microwave domain~\cite{ground1,enOM1,Lehnert,Fink}, is much larger than the mechanical frequency, and in this case electro-mechanical entanglement can be directly generated by adopting an appropriate microwave drive~\cite{DV07}. The entanglement generated in this work, however, uses a different mechanism and is of low-frequency resonators (both around 1 MHz), which implies its macroscopic quantum nature, and would allow us to test quantum theories at a more macroscopic level~\cite{Bell1,Bell2,Bell3}.

\section*{Acknowledgments}

We thank D.\ Vitali and N.\ Malossi for fruitful discussions and valuable comments on the manuscript, and M.\ Forsch, I.\ Marinkovi\'c, G.\ Steele, R.\ Stockill, and A.\ Wallucks for useful discussions on LC circuits. This project was supported by the European Research Council (ERC StG Strong-Q, 676842) and by the Netherlands Organisation for Scientific Research (NWO/OCW), as part of the Frontiers
of Nanoscience program, as well as through a Vidi grant (680-47-541).

\section*{Appendix:  ``optical-spring" effect in the opto-electro-mechanical system}

Here we show how to derive the effective mechanical frequency in our strongly coupled tripartite system. We solve the QLEs \eqref{QLE2} in the frequency domain by taking Fourier transform of each equation and derive the effective mechanical susceptibility, through which we extract the effective mechanical frequency. 

Solving separately two quadrature equations for each mode, we obtain the solutions given in terms of the natural susceptibilities of the three subsystems, which are
\begin{widetext}
\begin{equation}\label{QLE3}
\begin{split}
\chi_c^{-1}(\omega) \, \delta X (\omega)&= G \Delta \delta x(\omega) + \! \sqrt{2 \kappa} \, \big[(\kappa -i \omega) X^{\rm in} (\omega) + \Delta Y^{\rm in} (\omega) \big],  \\
\chi_c^{-1}(\omega) \, \delta Y (\omega)&= (\kappa - i \omega) G \delta x(\omega) +  \!\sqrt{2 \kappa} \, \big[(\kappa - i \omega) Y^{\rm in} (\omega) - \Delta X^{\rm in} (\omega) \big],  \\
\chi_m^{-1}(\omega) \, \delta x (\omega)&= G \delta X (\omega) - g \delta q (\omega)  + \xi (\omega),   \\
\chi_{LC}^{-1}(\omega) \, \delta q (\omega)&= - g \delta x (\omega)  + \delta {\cal V} (\omega),   \\
\end{split}
\end{equation}
\end{widetext}
where $\chi_c(\omega)$, $\chi_m(\omega)$, and $\chi_{LC}(\omega)$ are the natural susceptibilities of the cavity, mechanical, and electrical modes, respectively, given by
\begin{equation}\label{3chi}
\begin{split}
\chi_c(\omega)&= \frac{1}{\Delta^2 + (\kappa - i \omega)^2 }    , \\
\chi_m(\omega)&= \frac{\omega_m}{\omega_m^2 - \omega^2 - i \gamma_m \omega }    , \\
\chi_{LC}(\omega)&= \frac{\omega_{LC}}{\omega_{LC}^2 - \omega^2 - i \gamma_{LC} \omega }    . \\
\end{split}
\end{equation}
The mutual interactions among the three modes lead to the modification of their natural susceptibilities, and thus yield effective mode frequencies, which are associated to the real part of the reciprocal of the susceptibilities. Inserting $\delta X (\omega)$ and $\delta q (\omega)$ in Eq.~\eqref{QLE3} into the equation of $\delta x (\omega)$, we obtain 
\begin{equation}\label{eqx}
\begin{split}
\chi_m^{\rm eff-1}(\omega) \, \delta x (\omega) =&  \chi_c(\omega)  G \! \sqrt{2 \kappa} \, \Big[(\kappa -i \omega) X^{\rm in} (\omega) + \Delta Y^{\rm in} (\omega) \Big]   \\
&+ \xi (\omega) - \chi_{LC}(\omega) g  \delta {\cal V} (\omega) ,   \\
\end{split}
\end{equation}
where $\chi_m^{\rm eff}(\omega)$ is the effective mechanical susceptibility, defined by
\begin{equation}
\chi_m^{\rm eff-1}(\omega) =  \chi_{mc}^{-1}(\omega) - g^2  \chi_{LC}(\omega) ,
\end{equation}
with 
\begin{equation}
\chi_{mc}^{-1}(\omega) =  \chi_{m}^{-1}(\omega) - G^2 \Delta \chi_c(\omega),
\end{equation}
where $\chi_{mc}(\omega)$ corresponds to the effective mechanical susceptibility in the presence of {\it only} the optomechanical interaction. From the real part of $\chi_m^{\rm eff-1}(\omega)$, we extract the effective mechanical frequency, where we can recognize the so-called ``optical-spring" effect which is accompanied by a further shift due to the electro-mechanical coupling, i.e.,
\begin{equation}
\begin{split}
&\omega_m^{\rm eff}(\omega) =  \\
& \Bigg[ \omega_m^2 - \frac{ G^2 \Delta \omega_m (\Delta^2 + \kappa^2 - \omega^2)}{ (\Delta^2 + \kappa^2 - \omega^2)^2 + 4 \kappa^2 \omega^2  }  - \frac{ g^2 \omega_{LC} \omega_m  (\omega_{LC}^2 -\omega^2 ) }{ (\omega_{LC}^2 -\omega^2 )^2 + \gamma_{LC}^2 \omega^2  }    \Bigg]^{\frac{1}{2}}.
\end{split}
\end{equation}  

\end{document}